# Fast and accurate extraction of ultra-high quality factor from cavity ring-down measurement

Yanping Yang[1,†], Shihan Liu[1,†], Yong Geng[1], Huashun Wen[2,*] and Heng Zhou[1,*]

[1]Key Lab of Optical Fiber Sensing and Communication Networks, School of Information and Communication Engineering, University of Electronic Science and Technology of China, Chengdu 611731, China
[2]State Key Laboratory on Integrated Optoelectronics, Institute of Semiconductors, Chinese Acacmy of Sciences, Beijing 100083, China
*Corresponding author: whs@semi.ac.cn; zhouheng@uestc.edu.cn †Those authors contributed equally.

**Cavity ring-down is an essential test to measure ultra-high quality factor (UHQ) optical cavities, which is, however, frequently misinterpreted due to lacking of a specified analysis guideline. Here we clarify the basic property of cavity ring down and present a step-by-step method that enables extraction of the overall quality factor, as well as the intrinsic loss and coupling state of UHQ cavities with better fidelity and simplicity than prior schemes. Our work can facilitate acurrate design and characterization of UHQ cavities for ultra-low noise lasers, high finesse reference cavities, and ultra-narrow optical filters.**

http://dx.doi.org/xxxxxx

Optical cavity is a fundamental component of modern optic and photonic systems, found ubiquitously in lasers[1,2], nonlinear optical devices[3,4], gyroscopes[5,6] and interferometers[7,8]. One of its essential parameters is the quality factor (Q-factor) defined as the ratio of the light energy stored in the cavity to the energy decayed during each roundtrip. Q-factor determines the spectral purity as well as the field enhancement of an optical cavity, so it is important to know accurately about the overall Q-factor $Q_T$, the intrinsic-loss induced $Q_i$ and out-coupling induced $Q_e$ so that to optimize target functions.

Conventionl Q-factor measurement can be categorized as *stationary* scheme and *dynamical* scheme [9-11]. *Stationary* scheme uses a continuous-wave (CW) laser to sample the steady-state power transmittance at densely chosen frequency detuning ($\delta$) with respect to the resonance (i.e., the laser dwell time at each $\delta$ is kept sufficiently longer than the cavity photon lifetime $\tau_\mathrm{p}$), producing a Lorentzian linewidth whose full-width-half-maximum (FWHM) equals to $\frac{\omega_0}{Q_T} = \frac{\omega_0}{Q_i} + \frac{\omega_0}{Q_e}$, namely, Q-factor can be obtained by recording the power transmission FWHM linewidth at certain cavity resonance frequency $\omega_0$. While being simple to implement, *stationary* scheme is inherently unable to reveal the exact values of $Q_i$ and $Q_e$ [9]. Moreover, for ultra-high quality factor (UHQ) cavity resonances (e.g., $Q_T$ exceeds $1\times10^9$) that simultaneously have super narrow linewidth (<200 kHz) and prolonged photon lifetime $\tau_\mathrm{p}$ (>1 μs), *stationary* scheme usually become incapable to measure Q-factor as it entails highly demanding laser frequency coherence and tuning linearity to produce clear Lorentzian lineshapes [11, 12].

*Dynamical* scheme, alternatively, scans laser frequency rapidly across the resonance before the intracavity field reached steady-state (allowing fast frequency scanning related to $\tau_\mathrm{p}$, thus suitable for UHQ cavities), and records the real-time power transmittance, which exhibits as a quasi-exponentially decaying wave-form called cavity ring-down (CRD) [13-14]. In addition, different with *stationary* scheme, *dynamical* CRD measurement is insensitive to the photo-thermal effect of the optical cavity undertest and the Q-factor can be determined without severe thermal distortion. In particular, *dynamical* scheme measures the cavity quality factor in the time domain by scanning a probe laser across the cavity resonance and quickly gating-off the laser power when it reaches the zero detuning point (i.e, resonance dip), then, the photon lifetime $\tau_\mathrm{p}$ and Q-factor can be inferred by fitting the decay rate of the intracavity light field [15]. However, this approach usually requires complex setup to ensure that the probe laser can be gating-off at the zero detuning point of the cavity resonance. Besides, a simpler CRD-based Q-factor measurement can be conducted in a couintous-wave fashion without gating off the laser field [13], but the analysis of the ringing curve requires more attentions. In this paper, we clarify the basic property of continuous-wave CRD measurement and present a specified algorithm that enables extraction of the $Q_T$, $Q_i$, $Q_e$ of UHQ cavities with superior acurrancy and simplicity than prior schemes.

Mathematically, the intracavity field $E_{cav}$ and output field $E_{out}$ from an optical cavity can be expressed as [9, 12]:

$$\frac{dE_{cav}(t)}{dt} = -\left[\left(\frac{\omega_0}{2Q_i} + \frac{\omega_0}{2Q_e}\right) + j\delta(t)\right]E_{cav}(t) + \sqrt{\frac{\omega_0}{Q_e}}E_{in}(t) \quad (1)$$

$$E_{out}(t) = -E_{in}(t) + \sqrt{\frac{\omega_0}{Q_e}}E_{cav}(t) \quad (2)$$

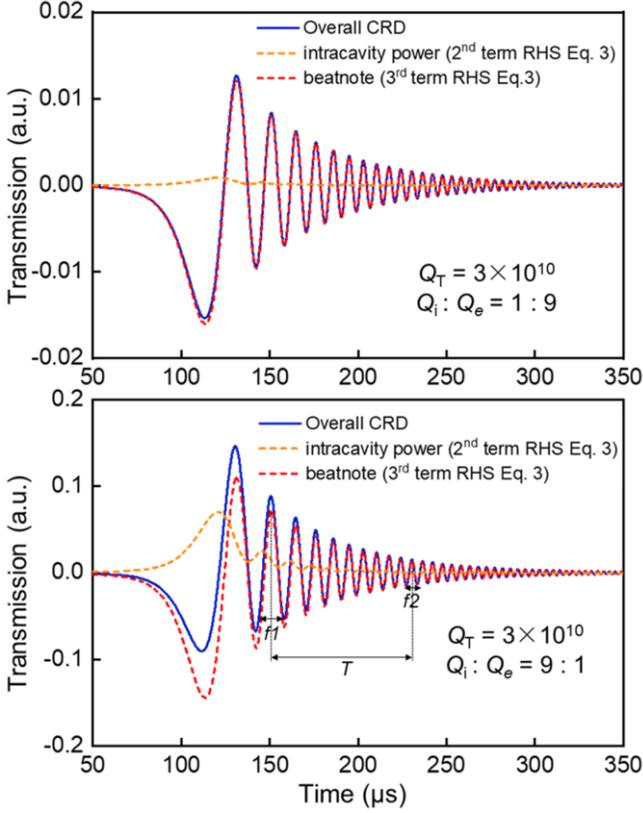

**Fig. 1** Theoretical calculated CRD with the same $Q_T$ but different $Q_i : Q_e$. The constant intput laser power (1st term RHS Eq. 3) is removed. It is seen that for over-coupled resonance (lower panel), the overall CRD dacay trajectory dose not reveal accurate $\tau_a$ nor $\tau_p$, while the rearmost oscillation waveform approximately predicts $\tau_a$.

$E_{in}$ is the incident field, $\delta = \omega_0 - \omega_p$ is the detuning between the resonance frequency $\omega_0$ and incident laser frequency $\omega_p$. CRD can be calculated by solving Eq. (1-2) with rapidly changing $\delta$ (mimicking frequency scanning) and recording the output power:

$$|E_{out}(t)|^2 = |E_{in}|^2 + \frac{\omega_0}{Q_{ext}}|E_{cav}|^2 - \sqrt{\frac{4\omega_0}{Q_{ext}}}|E_{in}E_{cav}| \qquad (3)$$

It is seen from Eq. (1-3) that CRD contains information of $Q_i$ and $Q_e$, which, however, can not be explicitly revealed by directly measureable CRD signatures (like the Lorentzian FWHM in *stationary scheme*), instead it requires perplexing fitting of $Q_i$, $Q_e$ and laser scanning speed(noted as *Vs* below) to reproduce the exact CRD [12], Importantly, we noticed that, frequently in prior literature, CRD measured from various UHQ cavities were incorrectly analyzed to extract $Q_i$ and $Q_e$ [16-21], due to lacking of a specified analysis guideline. It is therefore imperative to clarify the essential physical property of CRD and present a fast and accurate algorithm for $Q_i$ and $Q_e$ extraction.

In particular, the 1st term RHS of Eq. (3) is a constant representing the input laser power, the 2nd term is the out-coupled intracavity power, and the 3rd term is the heterodyne beatnote between the scanning input laser and intracavity field. Fig. 1 shows exemplified waveforms calculated from Eq. (1-3). Importantly, note that the majority oscillation waveform comes from the beatnote (3rd term RHS Eq. 3, proportional to $|E_{cav}|$), while the out-coupled intracavity power (2nd term RHS Eq. 3, proportional to $|E_{cav}|^2$) contribute only the minor part (depending on $Q_e$ [12]) within largely the first-half of the overall CRD decay trajectory. These fundamental features lead to two critical clarifications we want to emphasize:

**(1)** The beatnote decay time $\tau_a$ (indicated by the 3rd term RHS Eq. 3) is *twice* of the defined cavity photon lifetime $\tau_p$, then $Q_T = \omega_0 \tau_a / 2$. As seen in reference [16-22], $\tau_a$ was incorrectly deemed equal to the photon lifetime $\tau_p$, so the reported $Q_T$ therein were twice of the true values. For instance, in the publication [20], the CRD measured for silica microresonator (see Fig. 4 therein) was evidently misinterpreted, and the reported Q-factor was twice of the actual value.

**(2)** It is inaccurate to extract the CRD decay time $\tau_a$ starting from the foremost oscillation maximum [23], which contains the mixture of the 2nd and 3rd terms RHS Eq. (3). As seen in Fig. 1 low panel, especially for over-coupled resonance, starting from the foremost CRD generates notable error of $\tau_a$ and $\tau_p$, while the rearmost CRD waveform approximately predicts $\tau_a$ (i.e., the CRD curve coincides with the beatnote curve).

Indeed, when a CRD was experimentally measured, one can use Eq. (1-3) to numerically reproduce the measured CDR by abruptly fitting the values of $Q_i$ $Q_e$ and *Vs* (the changing rate of $\delta(t)$, and practically the laser frequency tunning speed), as done in previous studies [9, 11-12]. However, this undertaking could be perplexing and vulnerable to technique noise, especially when none of the three values were known in prior. Here, based on the above discussion, we present a detailed step-by-step recipe that is able to derive $Q_i$ and $Q_e$ from CRD rapidly and precisely.

**Step1:** Given a CRD curve, the rearmost ringing oscillations (assuming they have sufficient SNR) are used to roughly estimate the CRD decay time $\tau_a$, then an initial $Q_T$ can be evaluated as $Q_T = \omega_0 \tau_a / 2$.

**Step2:** Since CRD oscillations mainly come from the heterodyne beatnote between the intracavity field $E_{cav}$ and the frequency-scanning input field $E_{in}$ (3rd term RHS Eq. 3), we can approximately consider that $E_{cav}$ has fixed frequency as it contains laser fields whose frequency is limited within the narrow cavity resonance comparing with the rapidly-scanning frequency of $E_{in}$. So, $V_s$ can be estimated by the change rate of beating oscillation frequencies within the CRD curve, exprssed as: $V_s \sim (f_2 - f_1)/T$, here $f_2$ and $f_1$ represent the instantaneous frequency of the beating oscillations separated by a time interval $T$. Again, the rearmost oscillations with the CRD curve are preferable because they are less distorted by the intracavity power.

**Step3:** Adjust the ratio between $Q_i$ and $Q_e$ subject to the value of overall $Q_T$, to theoretically reproduce the experimentally measured CRD curve with as small as possible discrepancy [11].

**Step4:** Finely adjust the estimated values of $Q_T, Q_i, Q_e$ and $V_s$, to compensate technique errors and achieve the best match between measured and theoretical CRD curve, then finalize the parameters $Q_T, Q_i, Q_e$ and $V_s$ [9, 12].

The essential point of the above process is to extract useful and reliable information from the raw experimental CRD curve that can facilitate and expedite more accurate convergence of theoretical fitting of $Q_T, Q_i, Q_e$ and $V_s$ using theoretical formula Eq. 1-2 (comparing with abrupt fitting without any prior knowledge). We next conduct experimental validation of our method.

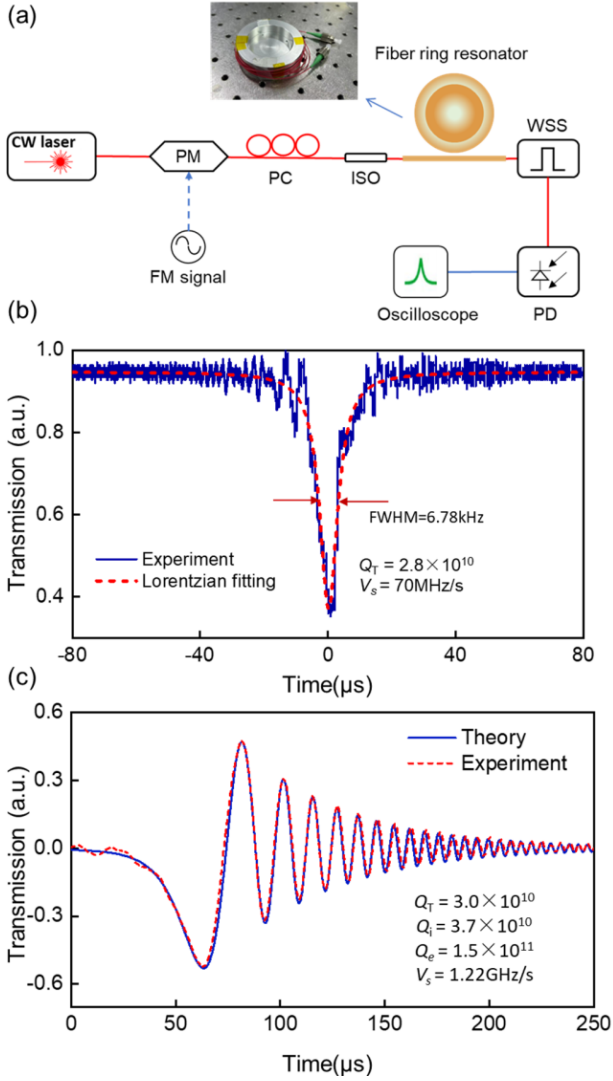

**Fig. 2** (a) Schemes to measure UHQ cavity using variable laser scanning speed enabled by frequency modulation. PC: polarization controller, PD: photodiode, PM: Phase modulator, ISO: optical isolator, WSS: wavelength selected switch. (b-c) Q-factor measurements of the same UHQ cavity under different laser scanning speed. (b) The Lorentzian lineshape obtained by *stationary* scheme. (c) The CRD curve measured by *dynamical* scheme.

First, Fig. 2 compares the Q-factor measurements of a UHQ fiber cavity based on both *stationary* and *dynamical* schemes. Since the laser modules available in our laboratory can only support linear frequency scanning with the slowest speed of several GHz/s, which is too fast to obtain smooth Lorentzian lineshapes in UHQ cavity for *stationary* measurement. To deal with this problem, we instead adopted external frequency modulation (FM) to scan the laser with flexibly controllable speed, the setup is shown in Fig. 2(a). An electro-optic phase modulator is driven by an RF source operating in linear FM mode. In particular, by changing the peak-to-peak voltage of the triangular FM driven signal (equal to the overall range of frequency chirping), the laser's actual frequency can be linearly scanning at various speed ranging from several MHz/s to several GHz/s. The FM laser is then sent through an all-pass fiber ring cavity, and the transmitted light is detected by an avalanche photodiode and recorded using a realtime oscilloscope.

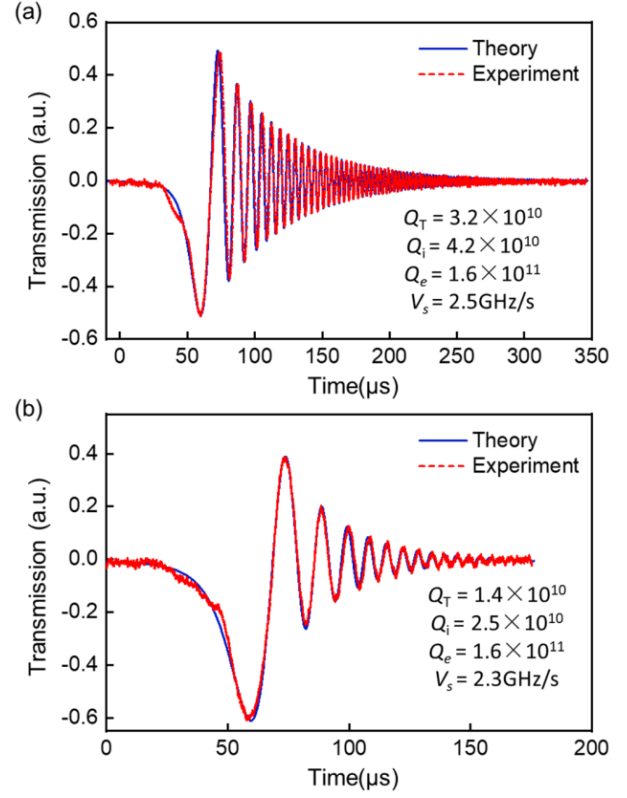

**Fig. 3.** Extraction of $Q_T$, $Q_i$, $Q_e$, from experimentally measured CRD, utilizing the proposed 4-step method. Solid blue line : the theoretical simulation waveform, red dashed line : experimental CRD of an UHQ cavity. (a) The CRD of an UHQ cavity with $Q_T$ =3.2×10¹⁰, $Q_i$ =4.2×10¹⁰, $Q_e$ =1.6×10¹¹. (b) The CRD of the same UHQ cavity after increasing intracavity loss and maintaining couple loss with $Q_T$ =1.4×10¹⁰, $Q_i$ =2.5×10¹⁰, $Q_e$=1.6×10¹¹.

Fig. 2(b) compares the results of both *stationary* and *dynamical* Q-factor measurements. When we configured the laser frequency scanning speed at 70 MHz/s, a Lorentzian lineshape were recorded with FWHM is 6.78 kHz, corresponding to the total Q-factor $Q_T = 2.8 \times 10^{10}$ ($\omega_0 = 2\pi \times 193.4$ THz). Then we increased the laser frequency scanning speed to about 1.22 GHz/s, and this time a CRD trace is recorded. Following our 4-step recipe described above, from the measured CRD curve we can obtain the following parameters: $\tau_a = 49$ μs; $Q_T = 3.0 \times 10^{10}$; $Q_i = 3.7 \times 10^{10}$; and $Q_e = 1.5 \times 10^{11}$. Comparing Fig. 2(b) and 2(c), we can conclude that *dynamical* measurement not only accurately extract the overall Q-factor $Q_T$ value (in consistency with *stationary* scheme), but also generate reliable values of $Q_i$, $Q_e$ and $V_s$, which cannot be derived by *stationary* scheme.

To further validate the accuracy of our method for *dynamical* Q-factor measurement, we also analyzed experimentally measured CRD curves from the same UHQ fiber ring cavity, but with slightly changed loss state. For this test the laser frequency scanning speed is set approximately at 2.5 GHz/s. Fig. 3(a) shows the original CRD of the device, implying $Q_T$ =3.2×10¹⁰, $Q_i = 4.2 \times 10^{10}$, $Q_e = 1.6 \times 10^{11}$. The CRD in Fig. 3(b) was measured by manually adding minor loss to the fiber ring, and our method accurately captures this feature and generates reduced $Q_i = 2.5 \times 10^{10}$, while maintaining unchanged $Q_e = 1.6 \times 10^{11}$. These results indicate the coupling loss is lower than the intrinsic loss in both cases, meaning that the UHQ cavity operates in the under-coupled state. Reliable results are repeated for multiple devices (not shown) that confirm the validity of our method. The minor discrepancy of $V_s$ between the two measurements is mainly caused by the imperfect repeatability of the laser frequency scanning speed.

In summary, as UHQ cavities are now ubiquitous in optics and photonics, rigorous elucidation of their fundamental property is imperative. We revisited and scrutinized the essential features of *dynamical* CRD measurement and present a step-by-step method that enables extraction of Q-factor with better fidelity and simplicity than prior schemes. Experimental verification of our method has been demonstrated with UHQ cavity ($Q_T \sim 3.2 \times 10^{10}$, $Q_e \sim 1.6 \times 10^{11}$, $Q_i \sim 4.2 \times 10^{10}$). Finally, we noticed that in the latest publication Optica 11(2), 176, 2024 [24], binary phase modulation of the incident laser was proposed to resolve $Q_i$ and $Q_e$. However, as phase modulation is equivalent to fast changing of instantaneous frequency, it is essentially a variant of CRD measurement that might be more difficult to obtain the precise values of $Q_i$ and $Q_e$, due to that the changing rate of instantaneous frequency becomes complicatedly time-dependent following the exact rising/falling edge of binary phase modulation pattern. Besides, for binary phase modulation, the input laser needs to be slowly tuned to the critical coupling point (akin to traditional laser-chopping-off CRD spectroscopy [13-14]), making the scheme unsuitable for UHQ cavities with super narrow resonances.

**Funding.** This work is supported by National Key Researchand Development Program of China (2021YFB2800602); National Natural Science Foundation of China (62001086, 62375043, U22A2086); Sichuan Province Science and Technology Support Program (2022YFSY0062, 23ZDYF3208); State Key Laboratory of Advanced Optical Communication Systems and Networks (2021GZKF010).

**Disclosures.** The authors declare no conflicts of interest.

**Data availability.** Data are available upon reasonable request.